\documentclass[conference]{IEEEtran}

\usepackage{amsmath,amssymb,amsfonts}
\usepackage{cite}
\usepackage{booktabs,tabularx,array}
\usepackage{graphicx}
\usepackage{balance}
\usepackage{algorithmic}
\usepackage{textcomp}
\usepackage{xcolor}

\def\BibTeX{{\rm B\kern-.05em{\sc i\kern-.025em b}\kern-.08em
    T\kern-.1667em\lower.7ex\hbox{E}\kern-.125emX}}

\title{Trade-offs in Decentralized Agentic AI Discovery Across the Compute Continuum}

\makeatletter
\newcommand{\linebreakand}{%
  \end{@IEEEauthorhalign}
  \hfill\mbox{}\par
  \mbox{}\hfill\begin{@IEEEauthorhalign}
}
\makeatother

\author{
  \IEEEauthorblockN{Patrizio Dazzi}
  \IEEEauthorblockA{\textit{Department of Computer Science} \\
    \textit{University of Pisa}\\
    Pisa, Italy \\
    patrizio.dazzi@unipi.it}
  \and
  \IEEEauthorblockN{Emanuele Carlini}
  \IEEEauthorblockA{\textit{Institute of Information Science and Technologies} \\
    \textit{National Research Council of Italy}\\
    Pisa, Italy \\
    emanuele.carlini@isti.cnr.it}
  \linebreakand 
  \IEEEauthorblockN{Matteo Mordacchini}
  \IEEEauthorblockA{\textit{Institute of Informatics and Telematics} \\
    \textit{National Research Council of Italy}\\
    Pisa, Italy \\
    matteo.mordacchini@iit.cnr.it}
  \and
  \IEEEauthorblockN{Saul Urso}
  \IEEEauthorblockA{\textit{Department of Computer Science} \\
    \textit{University of Pisa}\\
    Pisa, Italy \\
    saul.urso@phd.unipi.it}
}

\begin{document}
\raggedbottom
\maketitle

\begin{abstract}
Agentic systems deployed across the compute continuum need discovery mechanisms that remain effective across cloud, edge, and intermittently connected domains. In some emerging agentic architectures, decentralized discovery is already an active design direction, placing DHT-based lookup on the path toward agent directories. This paper studies the trade-offs among major structured-overlay families for agent discovery, comparing Chord, Pastry, and Kademlia as candidate indexing substrates within a shared control-plane framework. Using a benchmark subset centered on a 4096-node stationary comparison and a representative 4096-node churn benchmark, the paper characterizes how discovery reliability, startup behavior, and control-plane overhead vary across these overlays. The goal is to clarify the operating points they expose for agent discovery across edge-to-cloud environments.
\end{abstract}


\section{Introduction}

AI agents are moving from isolated assistants toward distributed application components spanning enterprise software, cloud platforms, and edge services. In that setting, discovery becomes a control-plane function: agents must locate suitable peers and services under scale, heterogeneous connectivity, and changing membership~\cite{dazzi2025iaia,carlini2023smartorc}. The agent directory is therefore not merely a registry but operational infrastructure whose latency, reliability, and communication overhead affect higher-level orchestration~\cite{muscariello2025agntcyads,muscariello2026adsdraft}. Industry forecasts reinforce the relevance of this problem by anticipating rapid growth of agent-enabled applications in the near term~\cite{gartner2025agenticapps,gartner2025agenticdecisions}.
AGNTCY provides a concrete motivation for this study. Its Agent Directory Service (ADS) already treats decentralized capability discovery as a first-class architectural component and places a Kademlia-oriented design on the implementation path~\cite{muscariello2025agntcyads,muscariello2026adsdraft}. The question addressed here is therefore not whether decentralized discovery is useful in principle, but which structured overlay family offers the most attractive operating point once decentralized discovery is assumed.

This question is particularly relevant in continuum settings, where discovery must remain usable across cloud tiers, edge domains, and intermittently connected participants. In such environments, misses are not merely lookup artifacts: they can propagate into delayed orchestration, repeated retries, or failed cross-tier coordination. For that reason, the suitability of a DHT family cannot be judged only by asymptotic lookup complexity, but by the reliability-cost regime it exposes under the workload and failure conditions of the target deployment. That is also why a family-level comparison is still useful despite the existence of many optimized variants and implementation-specific refinements. A design choice made at the family level constrains the later optimization space: whether the overlay relies primarily on structured maintenance, prefix-based routing, or aggressive parallel lookups already shapes the kind of control-plane budget the deployment is likely to face. The contribution of this paper is therefore not to rank every concrete implementation, but to clarify what kinds of operating regimes these canonical families tend to expose once they are repurposed as substrates for decentralized agent discovery.

\subsection{DHTs as Discovery Substrates}
Distributed hash tables (DHTs) remain one of the standard answers to decentralized lookup. Chord, Pastry, and Kademlia represent three canonical families with different routing and maintenance strategies~\cite{stoica2001chord,rowstron2001pastry,maymounkov2002kademlia}. At a high level, a DHT maps identifiers to nodes, stores descriptors according to that mapping, and routes lookups through overlay neighbors rather than through a central index. In the usual structured-overlay view, if $N$ is the number of participating nodes and $b$ is the effective routing fan-out, expected lookup cost in hops $h(N)$ and routing state per node $s(N)$ are commonly summarized as
\[
  h(N) = \Theta(\log_b N) = \Theta\left(\frac{\log N}{\log b}\right),
\]
\[
  s(N) = \Theta(\log_b N).
\]
These asymptotic forms intentionally hide the constants introduced by routing structure, replication, and maintenance. For an agent directory, however, those constants are often exactly what matters, because they determine how much control-plane work is required to make discovery usable under realistic deployment conditions.
We compare Chord, Pastry, and Kademlia as alternative substrates for decentralized agent discovery, focusing on how their routing, maintenance, and state-management choices translate into different reliability-overhead operating points.

\subsection{Scope and Contribution}
The evaluation is framed around the trade-off exposed to operators. Discovery success and recall measure effectiveness, messages/query captures control-plane cost, and P95 query latency provides a compact routing-efficiency signal. Two benchmark slices are used: a stationary comparison at $N=4096$ that separates immediate query admission from a $\log_2 N$ warmed start, and a representative churn benchmark at the same scale that tests whether those operating points survive under persistent departures and rejoins.
The contribution is empirical rather than architectural. We do not propose a new DHT or a new AGNTCY design. Instead, we place three established overlay families in a shared agent-discovery framework and show that they expose distinct control-plane regimes. Immediate queries reveal a substantial cold-start penalty for all three overlays. After a short warmup, all three recover full discovery success in the present workload, after which the main separation is cost and latency: Pastry is the cheapest operating point, Chord remains a higher-cost middle ground, and Kademlia pays the largest communication bill while achieving the lowest tail latency at the representative churn point.

\subsection{Paper Roadmap}
Section~\ref{sec:related} reviews prior work on agent discovery and DHT-based lookup substrates. Section~\ref{sec:approach} defines the workload abstraction, metrics, and experimental methodology. Section~\ref{sec:evaluation} reports the stationary and representative-churn results for Chord, Pastry, and Kademlia. Section~\ref{sec:conclusion} concludes.

\section{Related Work}\label{sec:related}

Three lines of work are especially relevant to this paper. The first concerns agentic infrastructures in which discovery is elevated from an application convenience to a systems component. AGNTCY and ADS provide the immediate architectural setting by treating the agent directory as decentralized control-plane infrastructure rather than as a local registry~\cite{muscariello2025agntcyads,muscariello2026adsdraft}. More broadly, the Internet of AI Agents perspective frames networked agents as distributed intelligence that must discover, coordinate, and interact across heterogeneous environments~\cite{dazzi2025iaia}. The contribution here sits inside that architectural direction, but focuses on a narrower design question: how different structured DHT families behave once agent discovery is already assumed to be decentralized.

The second line concerns service and resource management across the compute continuum. Prior work highlights how orchestration across cloud and edge domains is shaped by heterogeneity, dynamic resource visibility, and cross-tier coordination~\cite{carlini2023smartorc}. Some studies also examine adaptation and self-organization in decentralized edge settings, for example, through dynamic workload balancing and urgent resource allocation under changing operating conditions~\cite{carlini2025dynamic,besozzi2025spare,mordacchini2009challenges,baraglia2011group,baraglia2012godel,lulli2015distributed}. At a broader level, recent continuum literature emphasizes distributed orchestration and explicit service/resource visibility across heterogeneous cloud-edge deployments~\cite{dechouniotis2025continuum,donta2023continuum}. Recent service-discovery architectures for the edge-cloud continuum make a related point from the discovery side, emphasizing scalability, adaptability, and responsiveness under distributed service provisioning~\cite{farhoudi2025discovery}. Taken together, these works establish the operational relevance of discovery in continuum environments, but they do not ask the comparative question pursued here: when decentralized lookup is already on the table, which DHT family exposes which reliability-cost operating point for an agent-directory workload?

The third line concerns decentralized service discovery itself. Classic DHT systems such as Chord, Pastry, and Kademlia define the canonical structured-overlay families used for large-scale decentralized lookup~\cite{stoica2001chord,rowstron2001pastry,maymounkov2002kademlia}. A particularly relevant historical precedent is the comparative study of Li et al., which evaluated multiple DHT families under churn through a performance-versus-cost framework~\cite{li2005pvc}. Our study is inspired by that comparative perspective, but reinterprets it for agent discovery, where the overlay is evaluated as part of a decentralized control plane rather than as a pure key-based lookup service. The comparison in this paper is intentionally family-centric: it contrasts the design principles of Chord-, Pastry-, and Kademlia-like overlays rather than surveying the many later variants and hybrid schemes. More general distributed service-discovery work has long explored semantic forwarding, decentralized advertisement, and peer-to-peer matching without centralized registries~\cite{chakraborty2006distributed}. Other work studies richer discovery mechanisms on top of decentralized settings, for example, by incorporating trust, workflow semantics, or service-quality constraints~\cite{barclay2022trustable}. Those efforts contribute to discovery architectures or auxiliary mechanisms. By contrast, this paper keeps the discovery semantics fixed and compares the underlying lookup substrate directly, so that differences in success, control-plane overhead, and churn behavior can be attributed primarily to the DHT family rather than to changes in trust models, semantic matching, or orchestration logic.

\section{Approach}\label{sec:approach}

We model the agent directory as a decentralized indexing substrate, following the ADS view of capability discovery as a distinct control-plane function~\cite{muscariello2025agntcyads,muscariello2026adsdraft}. Candidate agents publish descriptors into the overlay, and querying agents perform capability-based lookups to retrieve candidate providers for higher-level selection or orchestration. Under this abstraction, the discovery layer is evaluated as a control-plane service rather than as an end-to-end application stack.
The comparison focuses on four operator-visible dimensions:
\begin{itemize}[nosep]
  \item discovery success and recall as effectiveness indicators;
  \item control-plane traffic as operational cost;
  \item P95 query latency as a compact routing-efficiency proxy;
  \item robustness under churn as a deployment-oriented stress condition.
\end{itemize}
All protocols run under the same AGNTCY-like workload abstraction. The shared catalog contains 50 skills, each agent advertises exactly one skill, and all runs target the same lookup objective (\texttt{skill\_05}) so that protocol differences are not biased by changing demand patterns. A lookup is counted as successful if it returns at least one published descriptor matching the target skill under the shared query rules. This keeps the comparison aligned with whether the directory can surface a usable provider, rather than with downstream ranking heuristics.

These metrics are intentionally service-oriented. Discovery success captures the most immediate operator question: whether the control plane returns at least one usable match. Recall complements success by asking whether relevant descriptors remain retrievable as the overlay evolves. Messages/query makes the communication bill explicit, which matters because any gain in robustness may be paid for with additional control-plane traffic. P95 latency is included as a compact proxy for routing efficiency and tail behavior rather than as an end in itself.

\subsection{Protocol and Workload Alignment}
The query process is also held fixed across protocols. Queries originate from internal nodes and run under concurrent scheduling, so query-resolution messages and maintenance traffic share the network. For the stationary benchmark, we compare two startup regimes at $N=4096$: an \emph{immediate} regime in which discovery requests are issued from the start of the run, and a \emph{warmed} regime in which requests are admitted only after bootstrap plus a warmup of $\log_2 N$. The churn benchmark uses a \emph{warmup}-only regime with the same $\log_2 N$ delay. Stationary runs use a query rate of 0.125 with a 40-unit workload horizon; churn runs use a query rate of 0.1 with a 60-unit horizon. All three protocols use the same replication baseline ($r=3$), publish TTL (60), and republish period (20), so that the main source of variation is the overlay family rather than protocol-specific publication policy.
The overlays themselves represent distinct design choices within the DHT family. Chord emphasizes orderly key-space responsibility and stabilization, Pastry combines prefix-based routing with structured neighbor sets, and Kademlia relies on XOR-distance routing and parallel lookups~\cite{stoica2001chord,rowstron2001pastry,maymounkov2002kademlia}. For an agent directory, these choices matter operationally: a discovery miss becomes a retry, a delayed orchestration step, or a failed candidate selection, while additional robustness may require substantially more control-plane traffic.

Because queries and maintenance overlap in time, the stationary analysis reports two complementary cost views. \emph{Observed cost} counts all messages seen during a discovery window and therefore captures the control-plane load experienced by an operator during live operation. \emph{Query-only cost} counts structured GET messages/query and isolates lookup traffic from concurrent publication and republish activity. Reporting both makes it possible to separate cold-start interference from the intrinsic routing cost of the lookup itself.

\subsection{Benchmarks}
The two benchmarks answer complementary questions at the same scale. The Stationary Comparison at Scale keeps the network and overlay aligned at $N=4096$, disables churn, and raises repetitions to 10 in order to obtain a steadier large-scale baseline while explicitly contrasting immediate versus warmed query admission. The Representative Churn Benchmark keeps the same protocol set at $N=4096$, enables exponential session churn with same-ID rejoins, and evaluates a single operating point with session mean 100 and downtime mean 30. This second benchmark is intentionally narrower than a full sweep: it is meant to test whether the stationary ranking survives at full scale once persistent departures and rejoins are introduced, rather than to map the entire churn surface.
To keep that operating point computationally realistic, churn runs introduce a publish-spread window equal to the republish period. This does not alter lookup semantics or routing logic; it simply avoids simulator-induced global republish bursts in which thousands of nodes republish at the same instant. Table~\ref{tab:cloud-benchmark-params} summarizes the effective parameters of the two scenarios as used in the paper.

\begin{table}[t]
\centering
\caption{Benchmark parameters.}
\label{tab:cloud-benchmark-params}
\small
\begin{tabularx}{\linewidth}{@{}l>{\centering\arraybackslash}X>{\centering\arraybackslash}X@{}}
\toprule
\textbf{Setting} & \textbf{Stationary} & \textbf{Churn} \\
\midrule
Nodes & \multicolumn{2}{c}{4096} \\
\midrule
Repetitions & 10 & 1 \\
\midrule
Workload horizon & 40 & 60 \\
\midrule
Startup regime & immediate or bootstrap $+$ $\log_2 N$ & warmup-only ($\log_2 N$) \\
\midrule
Churn & disabled & exponential sessions, same-ID rejoin \\
\midrule
Session mean & -- & 100 \\
\midrule
Downtime mean & -- & 30 \\
\midrule
Query rate & 0.125 & 0.1 \\
\midrule
Publish spread & 0 & 20 \\
\midrule
Target & \multicolumn{2}{c}{\texttt{skill\_05}} \\
\midrule
Top-k & \multicolumn{2}{c}{1} \\
\midrule
Topology & \multicolumn{2}{c}{Erdos-Renyi, avg. degree 6} \\
\midrule
Latency / loss & \multicolumn{2}{c}{1.0 / 0.0} \\
\midrule
Replication & \multicolumn{2}{c}{3} \\
\midrule
TTL / republish & \multicolumn{2}{c}{60 / 20} \\
\midrule
Catalog & \multicolumn{2}{c}{50 skills; 1 skill per agent} \\
\bottomrule
\end{tabularx}
\end{table}
Methodologically, the analysis uses aggregate service-level metrics that map directly to operator concerns. Mean discovery success is the primary effectiveness signal; recall captures whether relevant descriptors remain retrievable; observed messages/query is the main communication-cost proxy; query-only GET messages/query isolates lookup traffic; and P95 latency summarizes routing efficiency. The resulting setup is deliberately narrow but highly controlled: query objective, publication semantics, replication level, and timing parameters are aligned across overlays so that the observed differences can be interpreted primarily as substrate behavior. This choice deliberately sacrifices workload breadth in exchange for interpretability. In particular, it avoids confounding the overlay comparison with protocol-specific publish strategies, target-selection policies, or heterogeneous semantic-matching rules. The benchmark subset should therefore be read as a comparative probe of discovery substrates, not as an attempt to exhaust the full AGNTCY workload design space. Its value lies in making the reliability-cost structure of the three overlays visible under common assumptions before richer application semantics are layered on top.

\section{Evaluation}\label{sec:evaluation}

\begin{figure}[t]
  \centering
  \includegraphics[width=\linewidth]{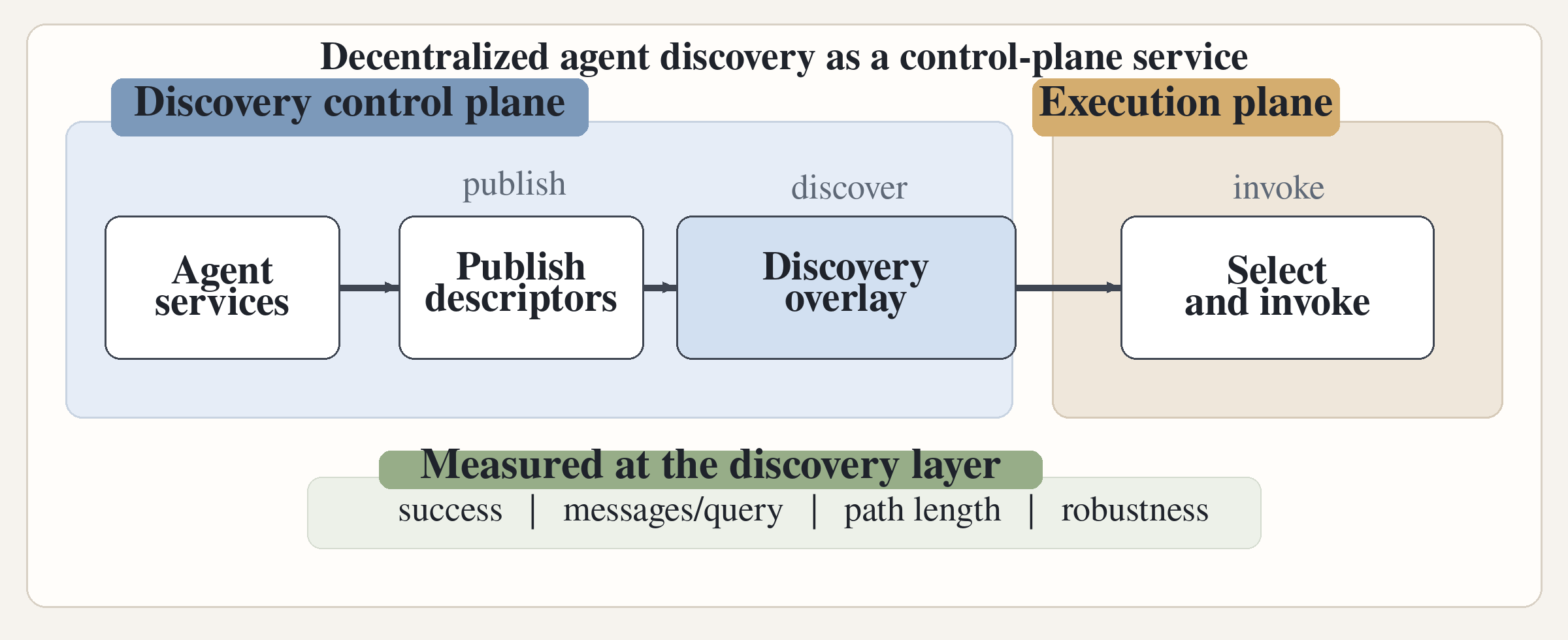}
  \caption{Conceptual workflow of decentralized agent discovery. Discovery is treated as a distinct control-plane layer: services publish descriptors into the overlay, queries retrieve candidate sets, and only then does orchestration proceed to invocation.}
  \label{fig:cloud-concept}
\end{figure}

\subsection{Benchmark Scope}
The evaluation reports results from a focused benchmark subset centered on two scenarios. The Stationary Comparison at Scale provides the large-scale baseline and explicitly separates cold-start behavior from post-bootstrap behavior by comparing immediate query admission against a $\log_2 N$ warmed start. The Representative Churn Benchmark keeps the same scale at $N=4096$ and tests a single churn operating point, asking whether the stationary ranking survives once departures and rejoins become persistent. Together, the two benchmarks answer the deployment question at the center of the paper: which design is cheaper to operate, which is more reliable, and whether churn changes the comparative picture enough to matter operationally.

\subsection{Stationary Comparison at Scale}
The full-scale stationary benchmark is best read as a startup comparison rather than as a single steady-state snapshot. At $N=4096$, issuing queries immediately yields only partial success: weighted mean success is 0.60 for Pastry, 0.62 for Chord, and 0.64 for Kademlia. Tail latency is also much worse in that regime, with P95 near 15.0 for Pastry, 30.0 for Chord, and 27.2 for Kademlia. Observed communication cost is correspondingly high, at about 10.2k messages/query for Pastry, 23.7k for Chord, and 70.7k for Kademlia. In other words, cold start is not a small perturbation at this scale; it changes both effectiveness and cost.
A short warmed start already changes the picture completely. After bootstrap plus a $\log_2 N$ warmup, all three overlays reach a weighted mean success of 1.0 and recall of 0.0122. P95 latency drops to 4.0 for Pastry, 7.0 for Chord, and 5.85 for Kademlia, while observed messages/query falls to about 3.6k, 7.5k, and 12.2k, respectively. Once startup interference is removed, correctness no longer separates the overlays; cost and latency do, with Pastry the cheapest operating point, Chord the middle ground, and Kademlia the most expensive.
The message disaggregation is important for interpreting these differences. Query-only GET cost stays comparatively small: in the immediate regime it is about 26.8 messages/query for Pastry, 63.7 for Chord, and 176.8 for Kademlia; after the $\log_2 N$ warmup it drops to 8.1, 14.5, and 16.7, respectively. The much larger observed overhead is therefore driven mainly by overlapping PUT traffic from publication and republish activity rather than by lookup routing alone. This distinction matters for reading cold-start results correctly: the immediate regime measures a live control plane during bootstrap, not just isolated lookup cost.

\begin{figure}[t]
  \centering
  \includegraphics[width=\linewidth]{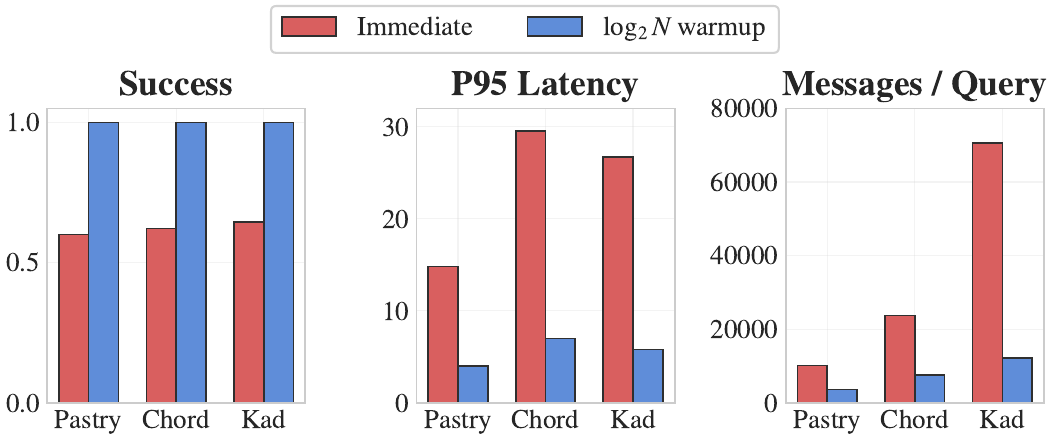}
  \caption{Stationary regimes at $N=4096$. A short $\log_2 N$ warmup removes most of the cold-start penalty in both latency and observed message cost.}
  \label{fig:cloud-stationary}
\end{figure}

\begin{table}[t]
\centering
\caption{Stationary regimes at $N=4096$. Entries are success / P95 / observed msgs/q / GET msgs/q.}
\label{tab:cloud-stationary-regimes}
\small
\begin{tabularx}{\linewidth}{@{}l>{\centering\arraybackslash}X>{\centering\arraybackslash}X@{}}
\toprule
\textbf{Protocol} & \textbf{Immediate} & \textbf{$\log_2 N$ warmup} \\
\midrule
Pastry & 0.60 / 15.0 / 10.2k / 26.8 & 1.00 / 4.0 / 3.6k / 8.1 \\
Chord & 0.62 / 30.0 / 23.7k / 63.7 & 1.00 / 7.0 / 7.5k / 14.5 \\
Kademlia & 0.64 / 27.2 / 70.7k / 176.8 & 1.00 / 5.9 / 12.2k / 16.7 \\
\bottomrule
\end{tabularx}
\end{table}

\subsection{Representative Churn Benchmark}
At the representative churn operating point ($N=4096$, session mean 100, downtime mean 30), all three overlays retain success 1.0 and precision 1.0 over the five-query workload. Recall is nearly identical as well, at 0.01525 for Pastry, 0.01533 for Chord, and 0.01484 for Kademlia. In other words, once the directory is given a short warmup, moderate persistent churn at full scale does not separate the overlays on correctness in this simplified workload.
The separation remains visible in latency and traffic cost. Pastry delivers P95 latency 5.0 with about 5.84k observed messages/query. Chord raises the traffic bill to 11.85k messages/query and P95 to 8.0. Kademlia reaches the lowest tail latency at 4.6, but still costs about 11.50k messages/query, close to Chord and roughly twice Pastry. This is the same qualitative trade-off seen in the warmed stationary regime: Pastry remains the cheapest operating point, while Kademlia spends substantially more control-plane traffic for a more aggressive lookup regime.
Because this benchmark samples a single full-scale churn point rather than a sweep, it should be read as a representative operating check, not as a full churn frontier. Its value is to show that the warmed stationary ranking is not an artifact of a no-churn setting: at $N=4096$, moderate churn still preserves full discovery success, and the main differences remain latency and operating cost rather than retrieval correctness.

\begin{figure}[thb]
  \centering
  \includegraphics[width=\linewidth]{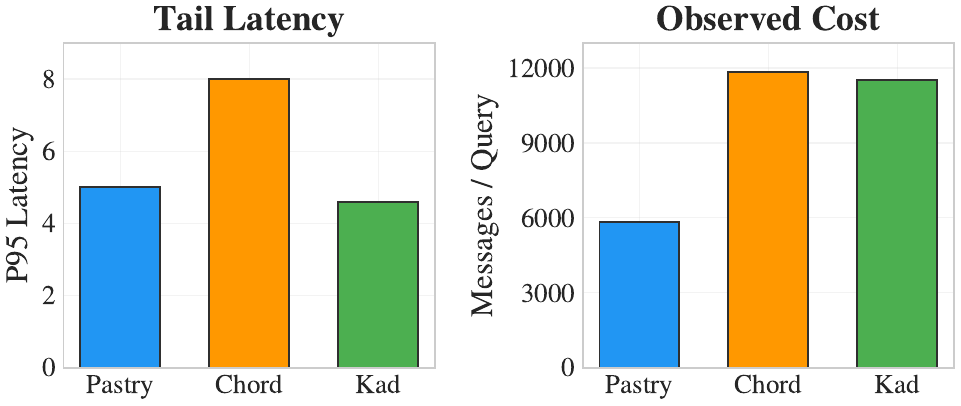}
  \caption{Representative churn operating point at $N=4096$. All protocols keep success at 1.0; the figure highlights the remaining latency and traffic trade-offs.}
  \label{fig:cloud-churn}
\end{figure}

\subsection{Cross-Benchmark Synthesis}
Taken together, the two benchmarks define distinct operating regimes rather than a single dominant substrate. Immediate queries at $N=4096$ expose cold-start behavior: discovery correctness is lower, tail latency is worse, and observed messages/query are inflated by concurrent publication traffic. A short $\log_2 N$ warmup is already sufficient to remove that startup penalty in the present stationary workload. Beyond that point, the family differences are easier to interpret: Pastry remains the cheapest operating point, Chord remains a higher-cost middle ground, and Kademlia trades a larger traffic bill for a somewhat more aggressive lookup profile.
This reading is closer to the actual deployment problem. If queries are admitted while the directory is still bootstrapping, what matters operationally is the full observed control-plane load, not just the routing cost of the lookup itself. If queries are admitted only after a short stabilization budget, the relevant distinction becomes the residual communication premium each overlay keeps once correctness has converged. At the representative churn point, that same distinction still holds: all three overlays preserve full success, so the practical choice is mainly about how much traffic the deployment is willing to spend for lower tail latency.

\subsection{Implications for AGNTCY Deployment}
These results matter most when read as guidance for how an agent directory is expected to fail in practice. In a deployment where discovery requests may arrive during cold start, the control plane should be evaluated on the full observed message load during the discovery window, because bootstrap-time publication traffic materially changes what the operator sees. In a deployment that can defer admission briefly, the comparison becomes cleaner: a $\log_2 N$ warmup is already enough to recover full success in the present stationary setting, after which Pastry minimizes cost, Chord offers a higher-cost middle ground, and Kademlia still pays a substantial communication premium.
The same comparison is also relevant from an architectural-governance perspective. AGNTCY and related agent platforms are still in a phase where directory-layer assumptions can harden into defaults. Once a family is adopted as the reference substrate, later engineering tends to optimize within that family rather than revisit the underlying lookup regime. A comparative result like the one presented here is therefore useful not only for immediate implementation tuning, but also for preventing a premature collapse of the design space onto a single canonical choice. In that sense, the value of the paper is partly methodological: it keeps the directory discussion open at the point where agent ecosystems are still deciding what kind of decentralized control plane they actually want to operate.

\begin{table}[t]
\centering
\caption{Observed operating points for AGNTCY discovery.}
\label{tab:cloud-guidance}
\small
\begin{tabularx}{\linewidth}{@{}>{\raggedright\arraybackslash}p{0.58\linewidth}X@{}}
\toprule
\textbf{Observed focus} & \textbf{Operating point} \\
\midrule
Lowest post-bootstrap cost & Pastry \\
Lowest tail latency at representative 4k churn & Kademlia \\
Lowest traffic at representative 4k churn & Pastry \\
Higher-cost middle ground across warmed regimes & Chord \\
\bottomrule
\end{tabularx}
\end{table}

\section{Conclusion}\label{sec:conclusion}

Across the evaluated benchmark subset, the main result is a stable trade-off between startup behavior, reliability, and operational cost across DHT families used for agent discovery. At $N=4096$, issuing queries immediately exposes a clear cold-start penalty for all three overlays: success drops below 1, latency rises, and observed per-query overhead is inflated by overlapping publication traffic. A short bootstrap-plus-warmup budget of $\log_2 N$ is already sufficient to remove that penalty in the present stationary workload. After that point, the steady comparison becomes simpler: Pastry occupies the lowest-cost operating point, Chord remains a higher-cost middle ground, and Kademlia still pays the largest communication bill.
Within the studied setup, the contribution is therefore to clarify control-plane operating regimes rather than to crown a universal winner. Different DHT families occupy different cost-reliability positions under a shared agent-discovery workload, and the 4096-node stationary comparison shows that large deployments do not collapse the comparison into a single outcome; instead, they make startup effects visible and make post-bootstrap cost differences easier to interpret. At the representative 4096-node churn point, all three overlays still preserve full discovery success, so the practical distinction remains mainly about how much traffic the deployment is willing to spend for lower tail latency.

At the same time, the paper should be read as a controlled step rather than as a final verdict. The current results are tied to the present protocol realizations, to the chosen publication and lookup semantics, and to an idealized network regime that suppresses congestion and packet-loss effects. Future work can therefore expand in at least three directions: broader workload diversity, richer failure models, and protocol-family tuning beyond the present baseline settings. Those extensions are unlikely to erase the trade-off structure identified here, but they could shift where the most attractive operating point lies for specific deployments. That is precisely why the comparative framing matters: it provides a stable way to reason about future agent-directory designs even as the surrounding ecosystem and implementation details continue to evolve. More broadly, the benchmark subset developed here can serve as a reusable reference point for later ADS-oriented studies. Even when future work introduces richer semantic matching, non-ideal networking, or different publication strategies, a compact stationary-plus-churn baseline remains valuable because it separates startup effects, lookup cost, and maintenance cost instead of conflating them.

\balance
\bibliographystyle{IEEEtran}
\bibliography{references}
\end{document}